\documentclass[lettersize,comsoc]{IEEEtran}
\usepackage{amsmath,amsfonts}
\usepackage{algorithmic}
\usepackage{algorithm}
\usepackage{array}
\usepackage[caption=false,font=normalsize,labelfont=sf,textfont=sf]{subfig}
\usepackage{textcomp}
\usepackage{stfloats}
\usepackage{url}
\usepackage{verbatim}
\usepackage{graphicx}
\usepackage{cite}
\usepackage{soul}
\usepackage{xcolor}
\usepackage{ctable}
\usepackage{upgreek}
\begin{document}

\title{A Perspective on the Impact of \\ Group Delay Dispersion in \\ Future Terahertz Wireless Systems}

\author{ Karl Strecker$^{*1}$, Sabit Ekin$^2$, and John F. O'Hara$^1$\\
\thanks{$^*$ karl.l.strecker@okstate.edu}
\thanks{$^1$ Karl Strecker and John O'Hara are with the School of Electrical and Computer Engineering at Oklahoma State University, Stillwater, Oklahoma.}
\thanks{$^2$ Sabit Ekin is with the Department of Engineering Technology \& Industrial Distribution at Texas A\&M University, College Station, Texas.}
\thanks{Revised manuscript submitted May 16, 2024.}
\thanks{This work has been submitted to the IEEE for possible publication.  Copyright may be transferred without notice, after which this version may no longer be accessible.}
}


\markboth{IEEE Communications Magazine,~Vol.~XX, No.~X, MONTH~YEAR}{}

\IEEEpubid{0000--0000/00\$00.00~\copyright~2023 IEEE}

\maketitle

\begin{abstract}
This article discusses the challenges and opportunities of managing group delay dispersion (GDD) and its relation to the performance standards of future sixth-generation (6G) wireless communication systems utilizing terahertz frequency waves. The unique susceptibilities of 6G systems to GDD are described, along with a quantitative description of the sources of GDD, including multipath, rough surface scattering, intelligent reflecting surfaces, and propagation through the atmosphere.  An experimental case-study is presented that confirms previous models quantifying the impact of atmospheric GDD. Several GDD manipulation strategies are presented illustrating their hindered effectiveness in the 6G context.  Conversely, some benefits of leveraging GDD to enhance 6G systems, such as improved security and simplified hardware, are also discussed.  Finally, a perspective on using photonic GDD control devices is provided, revealing quantitative benefits that may unburden existing equalization schemes. The article argues that GDD will uniquely and significantly impact some 6G systems, but that its careful consideration along with new mitigation strategies, including photonic devices, will help optimize system performance. The conclusion provides a perspective to guide future research in this area.
\end{abstract}


\section{Introduction}

\IEEEPARstart{T}{he} rapid evolution of connectivity is now compelling the emergence of sub-mmWave (terahertz) or 6G wireless systems.  The 6G vision includes many challenging requirements \cite{6G_vision}. The first  of these challenges is unprecedented data throughput (1 Tbps) \cite{6G_vision, Tbit_single_carriers}, which requires massive bandwidth and terahertz-frequency signals (0.1-1.0 THz). Second, 6G systems should feature integrated sensing and communication (ISAC) to collect real-time environmental data and channel state information (CSI).  This requires a critical evaluation of state-of-practice methods, such as the use of pilots to collect CSI, and waveforms, such as orthogonal frequency division multiplexing (OFDM).  Third, intelligent reflecting surfaces (IRSs) will enable spatially-aware systems to target individual users with highly directive beams. IRSs may enable communication around obstacles, reduce receiver complexity \cite{Tbit_single_carriers}, and improve network capacity significantly.  Finally, research suggests telecommunication technologies, including 6G, will consume 10\% of global energy usage by 2030 \cite{Power}. Energy efficiency is increasingly becoming part of the 6G vision, a fact future 6G standards must accommodate.\IEEEpubidadjcol

Now we draw attention to group delay dispersion (GDD).  We suggest it's a phenomenon affecting each of the areas listed above, though its impact in 6G may still not be fully appreciated. The GDD of a communication channel quantifies chromatic dispersion, i.e., the variations in propagation delay experienced by frequency components of a signal propagating through a channel. Practically, GDD reshapes signals in time and can cause intersymbol interference (ISI) when uncontrolled. Mathematically, GDD is the second derivative of phase with respect to frequency, $\frac{d^2\phi}{d\omega^2}$. Any channel element that imparts a nonlinear phase shift over frequency will exhibit GDD and reshape signals, regardless of the physical mechanism at work. For 6G, two related mechanisms are most relevant:  atmospheric dispersion and multipath, though others exist, as we elaborate below. 

Using this framework, the first half of this perspective article will elucidate how GDD impacts the aforementioned 6G challenges and why GDD in 6G links may be more impactful than previously assumed.  That is not to say GDD is only a burden; it is also a tool.  The second half of this article will explore potential innovation opportunities for managing and even leveraging GDD in 6G systems.

\section{The Challenge of GDD in 6G Systems}
As in previous wireless generations, GDD management \textit{will} be an issue in 6G systems. However, GDD in 6G imposes a triple-penalty in comparison. Extreme bandwidths introduce a greater range of frequencies over which group delay may vary, while high bit rates reduce symbol timing so less temporal spread results in ISI.  Combined, these mean the severity of GDD scales \textit{quadratically} with bandwidth, as discussed below.  On top of this, the channel itself is far more dispersive in the 6G bands for reasons we will elucidate.  Collectively, this suggests GDD will be more prevalent, more impactful, and harder to correct in 6G, compared to legacy systems.

\subsection{Sources of GDD}
A summary of the major sources of GDD in 6G links is presented in Table~\ref{tab:karl_label}.  These are detailed in the following text.  

\begin{table}[hb!]
    \caption{Summary of GDD sources and their properties.}
    \footnotesize
    \begin{tabular}{|m{0.08\textwidth}|m{0.155\textwidth}|m{0.16\textwidth}|m{0.025\textwidth}|}
        \hline
       {\bf Source} & {\bf Predictability} & {\bf Rate of change} & {\bf Ref} \\
       \specialrule{.2em}{0em}{0em} 
       Atmospheric resonances & Deterministic & Slow: minutes/hours & \cite{Fundamental} \\
       \hline 
       Multipath propgation & Deterministic with ray-tracing -- otherwise statistical & Fast ($10~\upmu$s) for mobile receivers or NLOS links - not appreciable for LOS & \cite{6G_vision} \\
       \hline
       Rough surfaces & Statistical & Static for stationary NLOS links - otherwise fast & \cite{Russ}\\
       \hline
       Structured surfaces & Deterministic & Equal to rate of surface modulation & \cite{Fahim}\\
       \hline
       
    \end{tabular}
    
    \label{tab:karl_label}
\end{table}

\subsubsection{Atmospheric Resonances}
\begin{figure}[!ht]
\centering
\includegraphics[width=3.5in]{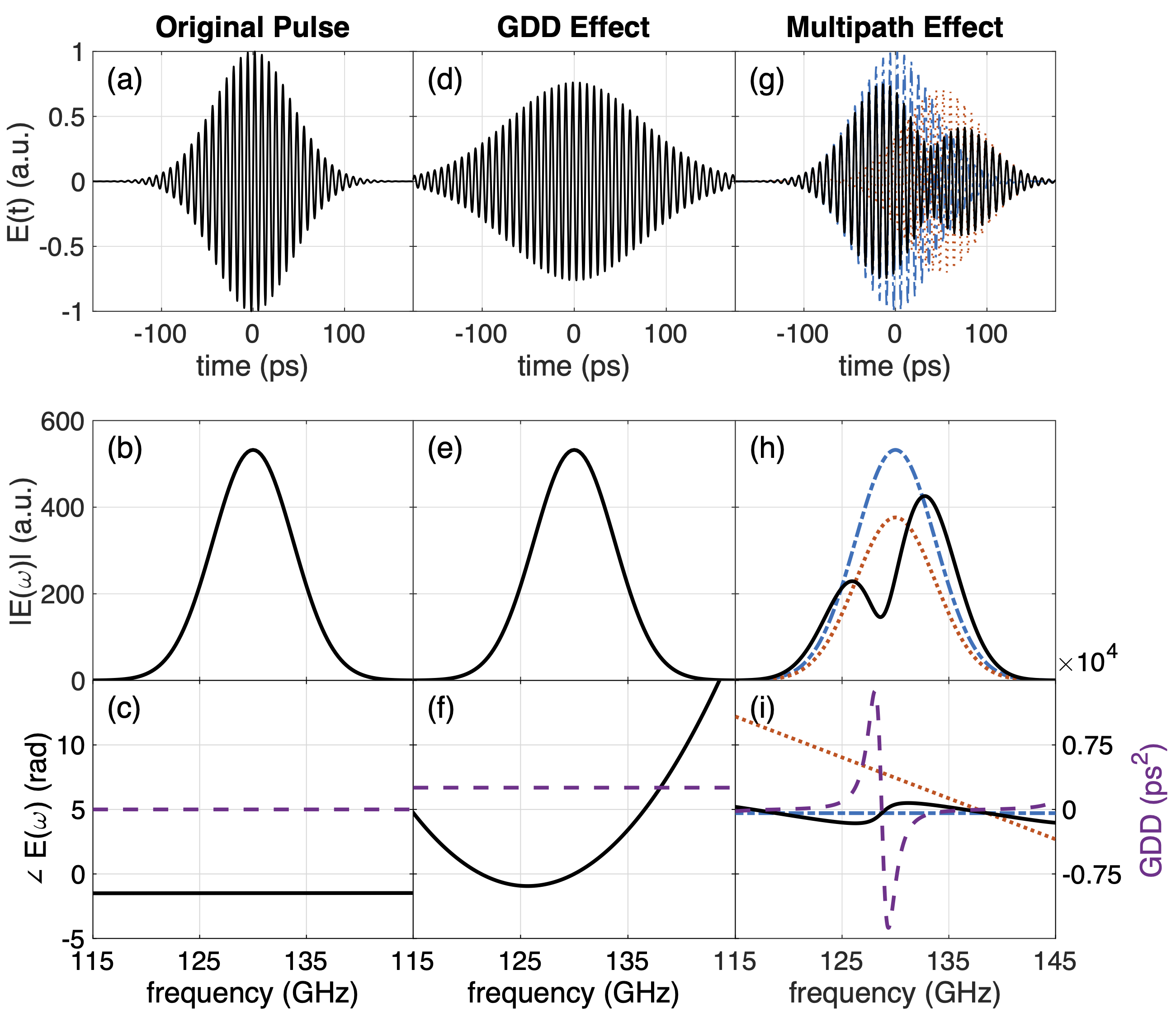}
\caption{Conceptual manifestations of GDD.  Left column shows a 100~ps pulse (a) at carrier $f_c=130$~GHz and its spectral amplitude (b) and phase (c).  Middle column shows the pulse stretched by GDD (d) resulting from a frequency-dependent channel refractivity. No change occurs to the spectral amplitude (e) but phase is now nonlinear (f).  Right column shows two 100~ps pulses, one delayed 50.5~ps (g).  Individual pulses are shown in dashed (red) or dot-dashed (blue); their sum is solid (black).  Fading due to multipath propagation is apparent in the summed signal spectrum (h).  Individual pulses have linear phase but their sum is nonlinear (i).  Quantitative GDD is in the bottom row, dashed (purple) curves, corresponding to the $y$-axis scale on the right.  For simplicity, true atmospheric refractivity was not employed in (f), however the GDD magnitude is realistic for multi-km backhaul links or operations near atmospheric resonances.}
\label{figGDD}
\end{figure}

\begin{figure}[!ht]
\centering
\includegraphics[width=3.5in]{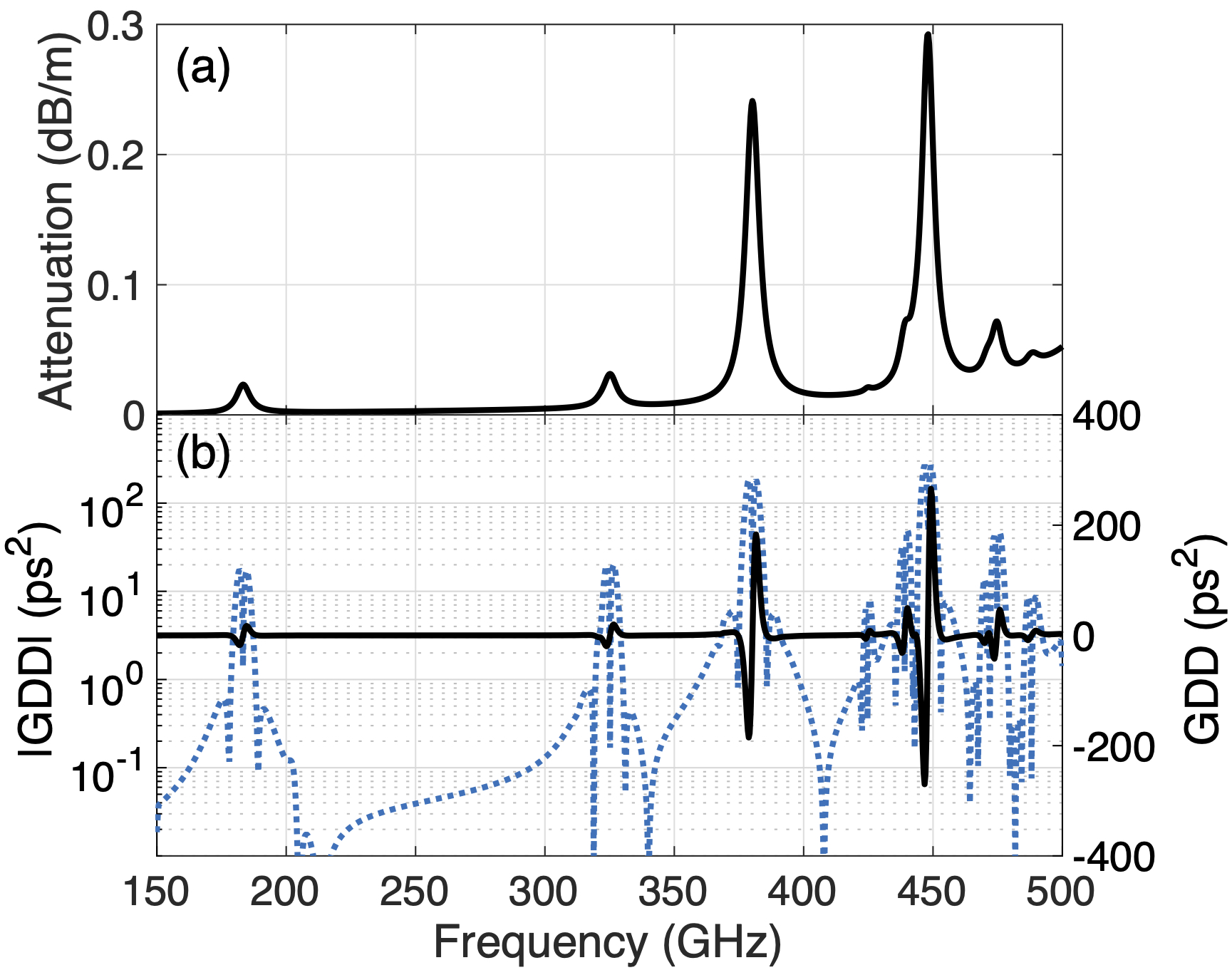}
\caption{Atmospheric attenuation (a) and GDD (b) at 29 $^{\circ}$C and 45\% relative humidity, normalized to a propagation distance of 1 meter. GDD is shown in log scale (blue dotted - left), and linear scale (black - right). }
\label{figatmos}
\end{figure}

GDD occurs when the cumulative phase characteristics (e.g. refractivity) of a material vary with frequency.  For example, chromatic dispersion is inherent in the glass comprising fiber optic cables.  A similar frequency-dependent refractivity occurs in 6G wireless systems due to atmospheric water vapor and diatomic oxygen absorption resonances.  Figure~\ref{figGDD} illustrates.  A broadband pulse is shown in the left column with its spectral amplitude and phase.  The center column shows the stretching effect of GDD in the time domain, commensurate with a nonlinear phase shift across its band.  With pure GDD, no frequency-varying effect on the spectral amplitude is evident. Existing models of the atmosphere's refractivity enable accurate calculation of atmospheric GDD, shown in Fig.~\ref{figatmos}. The plots show that GDD is highly elevated near the dispersive absorption resonances, but is relatively minor within the windows.  However, as previously shown \cite{Fundamental}, GDD is cumulative with propagation distance and can be significant within the windows when links span large bandwidths (tens of GHz) and kilometer-scale distances (i.e., high-capacity backhaul links). 

Additionally, 6G systems may not always operate in atmospheric windows. For security, it is desirable to operate on or near a vapor resonance to rapidly attenuate inadvertently scattered signals. This is also desirable in personal-area networking or small-cell communication, where a confined signal avoids interference among neighboring devices. Operating near vapor resonances might also prove necessary due to spectrum scarcity, as unfavorable propagation characteristics may be a valid trade-off in reducing spectrum crowding.

We have observed significant effects on the waveform when average bandwidth-integrated GDD \emph{magnitude} (related to group delay) exceeds the approximate duration of the symbol (for our work, $\approx 100$~ps).  This can take the form of a large GDD over a smaller bandwidth or vice versa.  For example, in the case of Fig.~\ref{figGDD}(f) this value is approximately $(2\pi\times 9\;{\rm GHz})(2533\;{\rm ps}^2)=143\;{\rm ps}$.  However, the required average GDD scales inversely with the square of bandwidth, i.e., high-bandwidth, shorter symbols require much less average GDD to observe significant distortion.

\begin{figure}[!ht]
\centering
\includegraphics[width=3.5in]{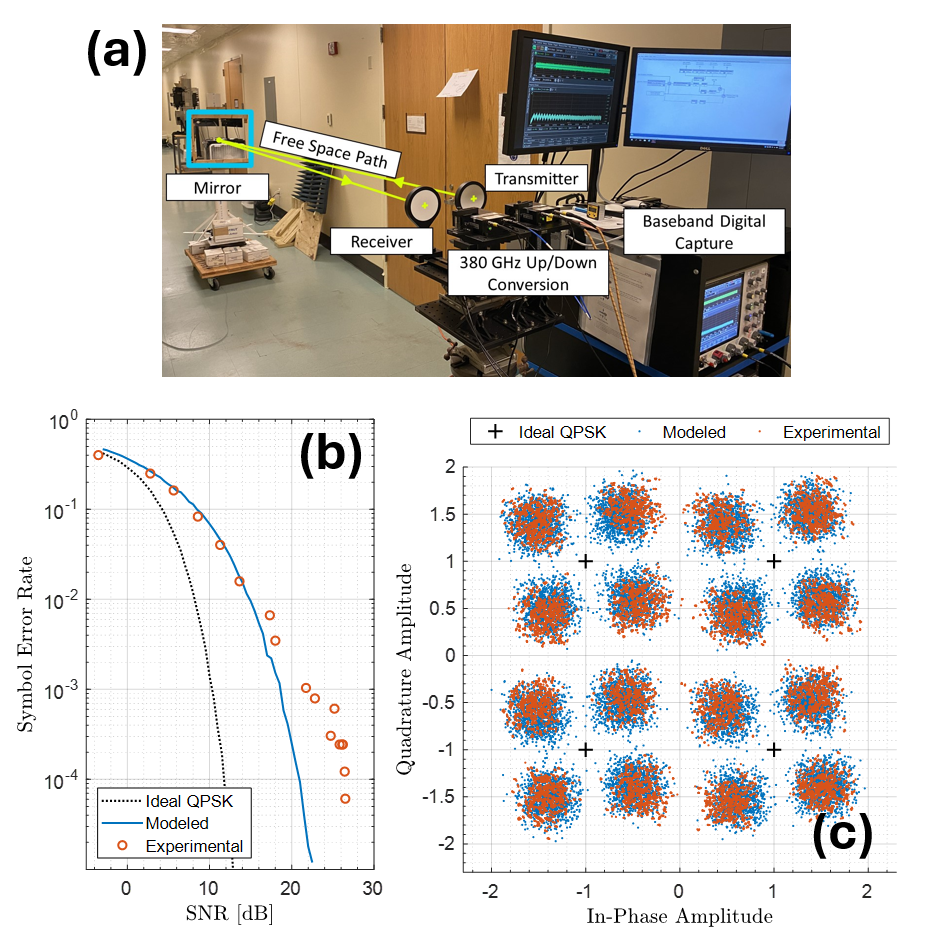}
\caption{Subfigure (a) is an annotated photograph of the 20 Gbit/s QPSK communication link at 380 GHz located in the climate-controlled hallway of the laboratory. In this image, the distance between the transmitter / receiver and the flat mirror (framed by the blue box) is only 3.8 meters to fit both the equipment and mirror within the image with acceptable clarity.  Subfigure (b) shows the experimental versus simulated symbol error rate.  Subfigure (c) shows the experimental versus simulated constellation.  In (c), both signals both have a SNR of 21.4 dB.}
\label{figexp}
\end{figure}

An empirical case-study readily illustrates these points.  Near water vapor lines, GDD-induced ISI is severe enough to be observed over short distances. To demonstrate this, the TWISTER system \cite{TWISTER} at Oklahoma State University (OSU) was configured as a single-carrier 20 Gbit/s QPSK wireless link at 380~GHz, transmitting 4096-bit pseudo-random data frames. The 30~meter wireless link was set up in a climate-controlled indoor laboratory at OSU, pictured in  Fig.~\ref{figexp}(a), with atmospheric conditions set to 29 $^{\circ}$C and 45\% relative humidity.  This resulted in a peak GDD of 5,500~ps$^2$. These conditions could realistically be observed in future 6G communication links.  The short channel induced significant ISI, resulting in the observed bit-error rate curve and constellation diagram shown in Fig.~\ref{figexp}(b) and (c).  The plots clearly reveal the strong impact on both BER and the constellation diagram structure.  They also demonstrate that atmospheric GDD can strongly affect 6G communications, even in otherwise ``well-behaved'' channels (short-range line-of-sight (LOS) links with no jitter or scintillation).  Additional case-studies in different bands, with different waveforms and much longer distances, may be observed in previous works \cite{Mary, Scirep}, which show GDD is an important consideration for optimizing the link.  It is impossible to detail the impact of GDD for every possible use case, but the empirical and simulation studies adequately show it cannot be ignored.
  
To accurately predict GDD's impact in more general scenarios a comprehensive model is necessary.  The aforementioned simulations were based on our published theoretical framework for quantifying the effect of atmospheric GDD on communication system performance.  Details can be found in \cite{Fundamental}, but are briefly summarized here.  The GDD model focuses on the phase and amplitude response of the channel during unguided signal propagation, abstracting out other details such as antenna gain, receiver sensitivity, etc.  The channel impulse response is computed by selecting all resonances of diatomic oxygen and water vapor between 0-5~THz from the HITRAN database and summing their collision-broadened response using molecular response theory in the frequency domain.  The atmospheric impulse response is then converted to a linear filter, which is used to model the LOS channel. GDD is deterministic and depends on the specific symbol sequences encountered, much like trellis states in maximum likelihood sequence estimation (MLSE) equalizers. Therefore, statistical effects of Gaussian noise are combined with deterministic sequence effects to provide the probability distribution of received symbols on a constellation diagram.  These result in BER predictions as a function of the signal-to-noise ratio. The modeled BER curves and constellation diagrams agree well with experimental results, as validated in Fig.~\ref{figexp}(b) and (c).

\subsubsection{Multipath Propagation}

Reflections, diffraction, and scattering can cause a signal to take multiple paths through the channel.  These paths experience varying delays, amplitudes, and phase shifts, which are summed at the receiver. In wideband (high data rate) communication scenarios with a short delay spread, the channel exhibits frequency-selective fading (FSF), causing certain frequency components to fade while others do not. This causes distortion and ISI in the time-domain, as illustrated in the right column of Fig.~\ref{figGDD}. To mitigate ISI caused by FSF, resource-intensive equalization strategies like maximum likelihood sequence estimation (MLSE) are employed, where simpler techniques like zero-forcing equalizers are generally insufficient. As illustrated in Fig.~\ref{figGDD}(i), multipath propagation also results in nonlinear phase shifts (GDD), which contributes to ISI differently from FSF. GDD is defined solely in terms of phase, while FSF impairs the signal via the fundamentally different mechanism of bandwidth reduction \cite{bandwidth_reduction}.  While FSF requires complex equalization techniques, GDD can be mitigated with simpler phase-shifting methods.  

\subsubsection{Surface Reflections}
Some 6G signals will be very directive and may be deliberately steered around obstacles by means of reflecting surfaces.  It is tempting to presume directive links will not experience multipath propagation, but this is not always true.  Surfaces with height variations exceeding ${\uplambda}/4$ are electromagnetically rough, forming dense scatterer clusters that significantly affect GDD in directive beams, as shown in prior work \cite{Russ}. The short wavelength of 6G systems means surfaces like textured drywall, rough concrete, and brick can cause measurable GDD.  For example, in \cite{Russ}, textured drywall with a mean surface height of $167\,{\rm \upmu m}$ and a height range of approximately $0-350\,{\rm\upmu m}$ was found to measurably degrade performance of a 20~Gbit/s communication link at 250~GHz. Furthermore, the severity of GDD was increased when multiple reflections occurred.

An alternative to natural reflectors, crucial to the 6G vision, is intelligent reflecting surfaces (IRSs), which include metamaterials, metasurfaces, and other engineered surfaces with specific electromagnetic responses. These devices will precisely steer 6G signals (often dynamically) and are envisioned for receiver tracking, multiplexing and passive modulation. However, given the highly phase-dispersive nature of typical metasurfaces, broadband IRSs present another possible source of undesired GDD \cite{Fahim} and may unintentionally reshape 6G waveforms as a byproduct of their primary function.  As a consequence of rough surface and IRS reflections, we can predict that non-line-of-sight (NLOS) links will thus suffer greater GDD on average than LOS links.

\subsection{The Challenges of Equalization}
Though GDD will occur in 6G wireless systems, the successful adoption of legacy wireless and fiber-optic mitigation strategies appears dubious. Here we highlight the potential shortcomings of commonly used approaches.

\subsubsection{Dispersion-tolerant waveforms are spectrally inefficient for 6G}
GDD reshapes time-domain waveforms and is therefore related to both modulation and pulse-shaping. In state-of-practice communication systems, OFDM and its variants are widely used. OFDM exhibits a high tolerance for temporal dispersion (i.e. GDD) by adding a guard interval before each symbol.  The cyclic prefix occupies this interval and mitigates ISI, but also reduces data rates and energy efficiency since it carries no additional information.

Furthermore, to be effective, the cyclic prefix must be longer than the channel's temporal dispersion.  According to \cite{Cprefix}, 6G would require drastically shorter OFDM symbol durations while keeping the guard interval unchanged, reducing spectral efficiency significantly. Challenges regarding frequency synchronization \cite{ref12} and peak-to-average-power-ratio (PAPR) \cite{ref11} further hinder implementing OFDM in terahertz systems.

Single-carrier and impulse radio modulation schemes lack many of the challenges of OFDM.  Accordingly, they are often proposed as candidates for 6G. However, these are more susceptible to GDD because they lack a guard interval between symbols.  Because 6G waveform design remains an open question \cite{ref11}, it seems premature to assume 6G waveforms will have GDD tolerance exceeding OFDM.

\subsubsection{Scaling digital equalizers will be challenging}
Currently, most GDD is mitigated using DSP, aka, electronic dispersion compensation (EDC).  This generally encompasses linear, nonlinear, and MLSE equalization, as well as digital pre-distortion.

EDC is currently used in both fiber optic channels and wireless systems.  However, 6G will combine the most challenging aspects of both these, exhibiting rapidly fluctuating channels with fiber-like bandwidths - a novel challenge.  With multipath propagation, wideband 6G wireless channels will regularly exhibit deep fading in the spectrum, which prohibits the effective use of low-complexity equalization techniques like zero-forcing that may be effective for fiber optics. 

Moreover, since wireless channels fluctuate over time, 6G equalizers will need to perform channel estimation and prediction. This is already done for 4G and 5G systems, but 6G will present greater challenges due to higher standards for mobile user velocity and the shorter wavelengths employed \cite{6G_vision}. For example, a 5G receiver operating at 28~GHz in a 108~km/h automobile  sees significant channel fluctuations in 0.18~ms (the half-wavelength travel time), whereas a 6G receiver operating at 250~GHz on a high-speed train (325 km/h) sees channel fluctuations in 6.65~$\upmu$s - over one order of magnitude faster. 

Channel estimation and some ISAC forms use periodic pilots (pre-arranged non-data sequences) for the receiver to update the CSI, which may be sent back to the transmitter for equalization. This process incurs training overhead, reducing spectral efficiency as equalizer weights are adjusted. Rapid 6G channel fluctuations require higher pilot density (over ten times greater), further decreasing spectral efficiency, similar to the challenge with OFDM cyclic prefixes.


To complicate equalization further, higher symbol rates, higher bandwidth, and more sources of GDD for 6G communications means the ``channel memory length'' (CML) could be orders of magnitude larger for some 6G systems, especially NLOS links.  This is problematic because MLSE, the most effective EDC method for fluctuating channels with deep fading, has a computational complexity that scales exponentially with CML \cite{MLSE}. There are adaptations to MLSE that reduce its complexity, but they do not remove the underlying exponential growth, making advanced equalization computationally prohibitive for many 6G links.  Even if the growth of computational power one day makes this problem tractable, it is likely that hardware capable of terahertz-frequency EDC will be specialized, complex, expensive, and power-hungry. Coupled with the increased power requirements of faster analog-to-digital and digital-to-analog converters to accommodate 6G Doppler shifts and multi-channel synchronization \cite{ref11, ref12}, this is not a welcome prospect.

For fixed 6G nodes (e.g., wireless backhaul) power consumption may not be crucial for performance, but remains concerning from a sustainability perspective as backhaul stations proliferate.  Moreover, most 6G end nodes (like aerial platforms, satellites, sensor nodes, and handheld mobile devices) will have limited processing power, memory, and energy budgets. Implementing advanced EDC architectures on these devices may be impractical, especially for battery-operated devices where energy efficiency is crucial. 

\begin{table}
    \caption{Summary of how GDD could detrimentally influence 6G systems}
    \begin{tabular}{|m{0.1\textwidth}|m{0.35\textwidth}|}
        \hline
       {\bf Issue} & {\bf 6G GDD complications} \\
       \specialrule{.2em}{0em}{0em} 
       Data rate & Short bit slot and higher bandwidth $\rightarrow$ more susceptibility to GDD.  Undesirable to increase pilot density \\
       \hline
       Carrier frequency & GDD due to atmospheric resonances unavoidable\\
       \hline 
       Waveform & GDD tolerant waveforms become spectrally inefficient\\
       \hline
       ISAC & GDD fluctuates more rapidly requiring more frequent CSI updates\\
       \hline
       Reflective surfaces &  More surfaces are electromagnetically rough and IRSs may be highly dispersive, increasing sources of GDD   \\
       \hline
       Energy efficiency & GDD may demand more DSP power to accommodate EDC \\
       \hline
       Cost and complexity & GDD more likely to span a wide CML, exponentially scaling EDC effort\\
       \hline
       
    \end{tabular}
    
    \label{tab:GDD links}
\end{table}

To summarize, GDD will certainly arise in 6G systems more frequently and with greater severity than in previous generations of wireless systems.  As discussed, it will thereby impact several other key system design factors, as summarized in Table~\ref{tab:GDD links}. GDD mitigation strategies such as OFDM cyclic prefixes and EDC architectures are undoubtedly effective for combating GDD due to multipath propagation in extant wireless systems, where they may represent the best overall tradeoff in terms of spectral efficiency, system complexity and system robustness. However, 6G systems impose greater demands in terms of throughput, channel fluctuation, available resources, and energy efficiency, making existing GDD mitigation strategies less attractive or possibly unviable.  

\section{The Opportunities of GDD in 6G Systems}
The second half of this perspective article encourages 6G architects to view GDD as a tool that can alleviate some of the challenges of the 6G vision.  It also suggests a new perspective on using novel photonic devices to mitigate the effects of undesired GDD.

\subsection{Using GDD to Enhance 6G Systems}
It is usually undesirable when symbols spread beyond their assigned time slots and overlap, but GDD could be used to mitigate the nuisance of PAPR in systems employing OFDM or pulse-based modulations. Waveform peaks could be intentionally flattened by introducing GDD before amplification and then perfectly restored thereafter, easing the burden on transceiver linearity.  This may prove to be a valuable step toward the potential implementation of OFDM in 6G. However, reversing the GDD in the signal post-amplification must be done photonically, not digitally.  This important point is discussed in the next section.

GDD may also enhance security.  By deliberately distorting the transmitted signal with severe and rapidly-changing GDD, spanning hundreds of symbol time slots, it could be made computationally prohibitive for an eavesdropper to digitally estimate and reverse GDD in real-time.  Again, MLSE techniques grow exponentially more expensive with increasing CML.  Knowing the GDD profile, the intended receiver could easily recover the signal with simple equalizers.  This approach may prevent man-in-the-middle eavesdropping, protecting any data that expires faster than an attacker could estimate and reverse the GDD.  

\subsection{Perspective on Photonic GDD Manipulation}

These considerations will lead to new innovations for GDD manipulation.   For example, the PAPR reduction scheme mentioned previously requires a photonic device in the channel to un-distort signals. This implies the use of a new class of GDD-manipulation technologies, namely, those relying on photonic dispersion compensation (PDC). 

PDC uses photonic methods to manipulate the phase of the electromagnetic wave directly, rather than performing computations on digital samples of the electric field. Although PDC has long been implemented in optics using dispersion compensating fiber (DCF), prism pairs, chirped mirrors, or fiber-based Bragg gratings, it has been either ignored in favor of DSP or rejected due to photonic devices' large size (commensurate with wavelength) and low modulation rate.

However, PDC becomes feasible in 6G where photonic devices are smaller and metasurface technology has advanced. Integrating PDC into multifunctional IRSs, for example, could offer seamless GDD correction alongside other functions in the 6G system (e.g. beamforming).  Furthermore, the same advances in IRS control that enable IRSs to perform spatio-temporal modulation will also enable IRSs to dynamically track changing GDD profiles.  Additionally, GDD is not \textit{always} stochastic; it can be deterministic and quasi-static (e.g., atmospheric GDD and GDD arising from IRSs).  This semi-static channel structure is already being leveraged for efficient channel coding schemes \cite{R1}, and could enable low-modulation-rate PDCs to mitigate channel GDD. If IRSs are designed with GDD management in mind, then they become a GDD solution, rather than a problem.

This idea can be extended further. PDC devices can be freely placed throughout the channel, while EDC is constrained to operate only at end nodes, suggesting the potential use of PDC IRSs as channel-integrated equalizers. This represents a paradigm shift regarding how equalization is performed. Currently, determining CSI and correcting impairments is the responsibility of individual end nodes.  In this vision, we make the environment \textit{itself} responsible for knowing CSI and for dynamically correcting any distortion it introduces.  This requires all PDC devices throughout the channel to predict the channel state for all users in the area.  This is no small challenge, but permanent PDC installations have an advantage over mobile end nodes since they can use high-resolution sensors (such as LIDARs and wide field-of-view cameras), coupled with machine learning networks trained to that specific location.  This can enable exceptionally accurate estimations of the channel state, and even \textit{predictive compensation} of channel impairments (such as GDD).

\begin{figure}[!ht]
\centering
\includegraphics[width=3.5in]{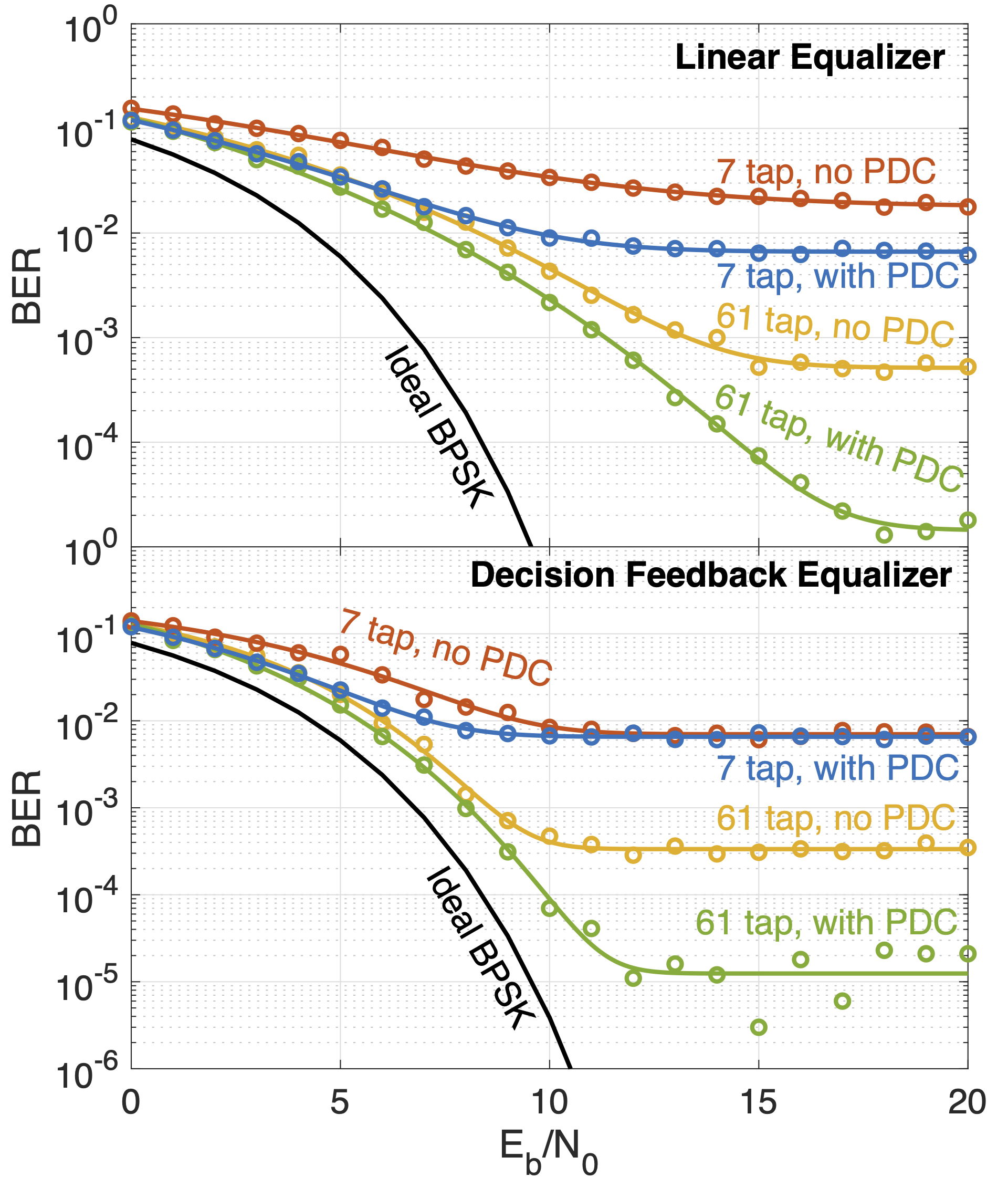}
\caption{Simulated performance of a direct 200~Gbit/s (0.2 Tbit/s) BPSK link through the atmosphere, contrasting links with and without PDC for both linear (top) and DFE (bottom) equalization strategies. The simulated link spans a distance of 100 meters. Note that the simulated PDC does not fully equalize the channel transfer function, but reverses GDD, only adjusting the phase of the signal.  For the DFE, the number of feed-forward and feedback taps are equal, and given by the tap number.}
\label{figEQs}
\end{figure}

Fig.~\ref{figEQs} provides an example of the simulated improvement in BER gained by using an integrated PDC to linearize the channel's phase before traditional EDC is performed.  In this example, a 0.2~Tbit/s BPSK link using 220-540~GHz with no multipath propagation experiences significant GDD and ISI because the signal spans multiple atmospheric water vapor resonances.  Fig.~\ref{figEQs} shows the result of equalization in 4 scenarios for both linear and decision feedback equalizers (DFE), namely, whether the equalizer fully spans the CML (61 tap cases) or not (7 tap cases), and whether or not PDC with was used to linearize the phase of the channel before EDC was performed.  In all cases, links using PDC prior to EDC performed better, even in a static channel case. This suggests that a significant fraction of the channel impairment is due to GDD and not attenuation.  It also demonstrates the benefit of having PDC devices with accurate CSI in the midst of the channel.  Additionally, Fig.~\ref{figEQs} suggests PDC could reduce the complexity while maintaining the performance of digital equalizers, or equivalently increase the performance of reduced-complexity equalizers. This is important, since 6G equalizers could otherwise become overly burdensome.  For example, ideal MLSE was not simulated in Fig.~\ref{figEQs} because the number of trellis states necessary to equalize the channel impulse response exceeded the memory allocation of the simulation computer.  

Channel-integrated PDC devices such as the one employed in Fig.~\ref{figEQs} would have myriad benefits beyond reducing the complexity of equalizers.  PDC would also reduce the necessary pilot density and the level of pre-distortion needed at transmitters by presenting a more stable channel (both discussed earlier as significant concerns for 6G). Having PDC placed within the channel itself is also economical, because it makes the most common devices (mobile end nodes) simpler and less costly.  It is also energy-efficient because photonic devices can be  very efficient themselves \cite{Scirep} and, importantly, consume resources in fixed installations rather than at resource-constrained nodes. Finally, PDC is powerful because it is not an either-or solution. PDC can operate independent of EDC, and hybrid configurations can be envisioned where PDC IRSs compensate for bulk GDD while simplified EDC architectures accommodate FSF and what little GDD remains.

To summarize, PDC traditionally hasn't been implemented because it didn't need to be.  However, the technology needed to adapt PDC to wireless applications is becoming ever more available, and 6G's unique needs present a compelling use-case. In our perspective, it seems essential to engage in research efforts toward the design of a new generation of sophisticated, smart, multi-functional photonic devices that can enhance security, reduce PAPR concerns, improve energy efficiency, and equalize rapidly fluctuating channels, working with other channel management and equalization technologies.

\section{Conclusion and Future Research Directions}
This article has presented a perspective on the growing challenges and opportunities of GDD in future 6G wireless systems. GDD arises from several sources, some practically unique to 6G.  
Experimental studies confirm the impact of GDD in real-world scenarios, and validate previous model predictions and simulations.  The discussions show that traditional methods of dealing with ISI caused by GDD may prove less effective for 6G systems. Tools such as cyclic prefixes and equalizers provide excellent design tradeoffs for legacy systems. However, employing the same tools for 6G presents significant and unique challenges.  The design, effectiveness, power efficiency, and complexity of tools required to manage GDD in 6G is possibly far from optimized. 

While future research should continue advancing legacy systems like EDCs for 6G, new avenues of GDD management, such as PDC, are ripe for exploration. Though multifunctional IRSs are not historically associated with PDC, these may now offer GDD management with superior performance, complexity, efficiency, and cost.  GDD may also be researched as an \emph{opportunity} toward improving security and reducing hardware complexity. Finally, hybrid equalization methods that combine PDC with EDC and ISAC may result in dynamic channels that unburden end nodes from costly DSP. Our perspective is that GDD deserves a much more careful examination as part of a comprehensive overall system design strategy.

\section*{Acknowledgments}
This material is based upon work supported in part by: 1) the National Science Foundation (NSF) Graduate Research Fellowship Grant No. 1746055 and by Grant Nos. 2018110 and 2238132; 2) the U.S. Department of Energy (DOE), Office of Science, Office of Advanced Scientific Computing Research under Award Number DE-SC0023023; 3) the National Aeronautics and Space Administration (NASA) via Grant No. 80NSSC22K0878.  Any opinions, findings, and conclusions or recommendations expressed in this material are those of the author(s) and do not necessarily reflect the views of the NSF, DOE, or NASA.

\newpage

\section{Biography Section}

\begin{IEEEbiography}[{\includegraphics[width=1in,height=1.25in,clip,keepaspectratio]{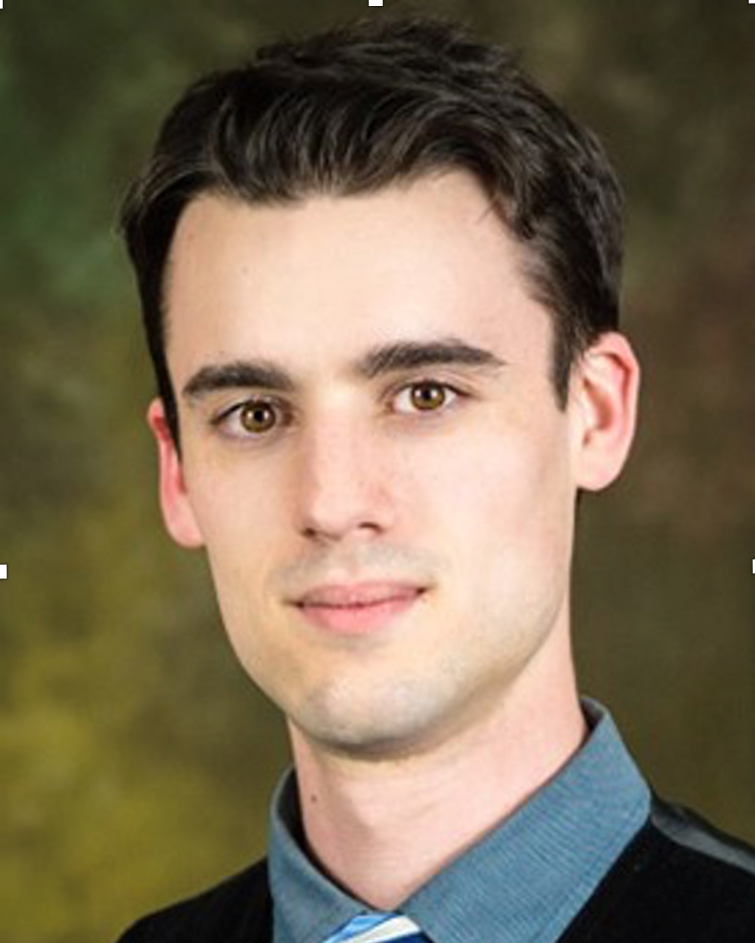}}]{Karl L. Strecker} received his Ph.D. degree in Electrical Engineering from Oklahoma State University in 2023. From 2018 to 2023, he worked as a student researcher in the Ultrafast Terahertz and Optoelectronics Laboratory at Oklahoma State University, where he now occupies a postdoctoral position. His research interests include wireless communication, terahertz channel characterization, and sub-millimeter-wave sensing.
\end{IEEEbiography}

\vspace{11pt}

\begin{IEEEbiography}[{\includegraphics[width=1in,height=1.25in,clip,keepaspectratio]{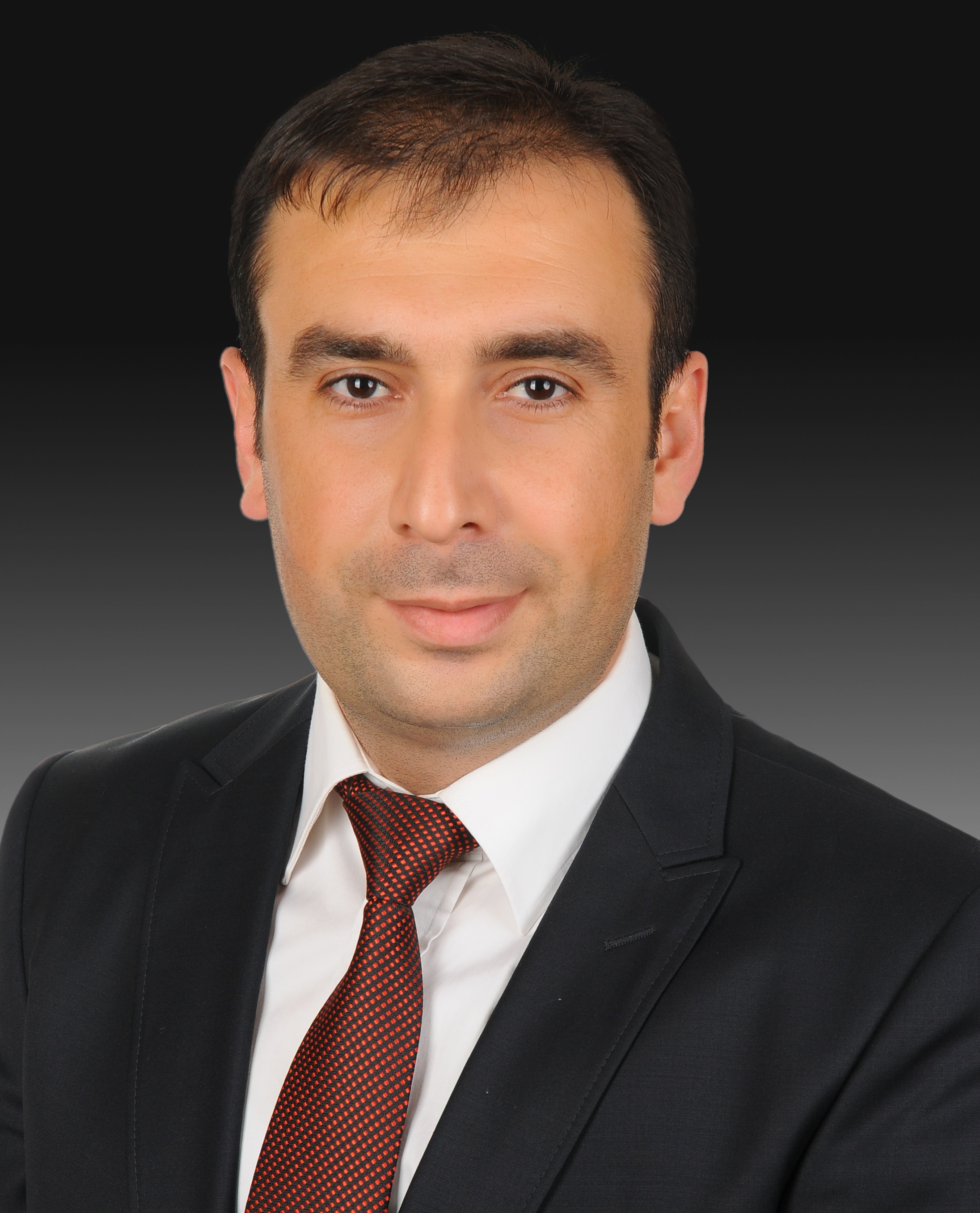}}]{Sabit Ekin} (SM'21) received his Ph.D. degree in Electrical and Computer Engineering from Texas A\&M University, College Station, TX, USA, in 2012. He has four years of industrial experience as a Senior Modem Systems Engineer at Qualcomm Inc. He is currently an Associate Professor of Engineering Technology and Electrical \& Computer Engineering at Texas A\&M University. His research interests include the design and analysis of wireless systems. 
\end{IEEEbiography}

\vspace{11pt}

\begin{IEEEbiography}[{\includegraphics[width=1in,height=1.25in,clip,keepaspectratio]{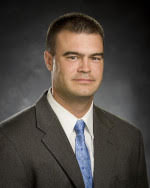}}]{John F. O'Hara} (M’05 - SM’19) received his Ph.D. (electrical engineering) from Oklahoma State University (OSU) in 2003.  From 2004-2011 he was with Los Alamos National Laboratory and worked on terahertz metamaterials and  technologies. In 2011, he founded a company, Wavetech, LLC specializing in automation and IoT devices. He is now the PSO/Albrecht Naeter Professor of Electrical and Computer Engineering at OSU and an NSF Career recipient. His  research interests are optical and terahertz wireless communications and metamaterials.
\end{IEEEbiography}

\vspace{11pt}

\vfill

\end{document}